\documentclass{article}

\usepackage{graphicx}
\usepackage[utf8]{inputenc}
\usepackage[T1]{fontenc}
\usepackage{cite}

\def\df{\partial}

\begin{document}

\title{New dark matter scenarios\\ with low energy photons and neutrinos}
\author{Igor Nikitin\\
Fraunhofer Institute for Algorithms and Scientific Computing\\
Schloss Birlinghoven, 53757 Sankt Augustin, Germany\\
\\
igor.nikitin@scai.fraunhofer.de
}
\date{}
\maketitle

\begin{abstract}
In this paper, a cosmological model is considered, in which dark matter is emitted by T-symmetric quasi-black holes distributed over galaxies. Low energy photons and neutrinos are taken as candidates for dark matter particles. Photon case can be closed due to the suppression of electromagnetic waves by interstellar/intergalactic medium. For neutrinos, two scenarios are considered. In the first, light neutrinos pair to bosonic quasiparticles that form massive halos around galaxies. In the second, heavier sterile neutrinos form massive halos without pairing. At the observables level, both scenarios turn out to be equivalent to the standard LambdaCDM model. At the same time, the considered model possesses additional sources -- quasi-black holes -- that can produce the required number of neutrinos in addition to the Big Bang. This opens up additional possibilities for cosmological constructions. The matching of the model with early cosmology and ongoing neutrino experiments is discussed. In addition, the possibility of solving general cosmological problems, including horizon, flatness and cosmological constant problems, as well as Hubble tension, within the framework of the considered model, is discussed.
\end{abstract}


\begin{figure}
\begin{center}
\includegraphics[width=0.7\textwidth]{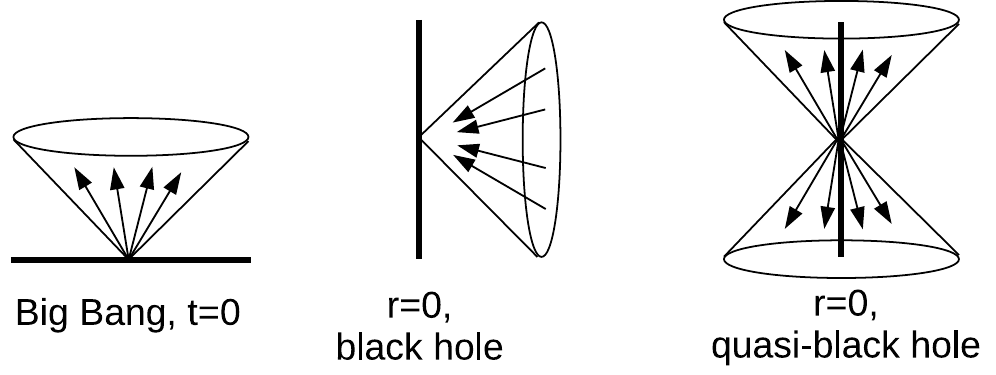}
\end{center}
\caption{The structure of light cones near singularities of various types.}\label{f1}
\end{figure}

\begin{figure}
\begin{center}
\includegraphics[width=0.7\textwidth]{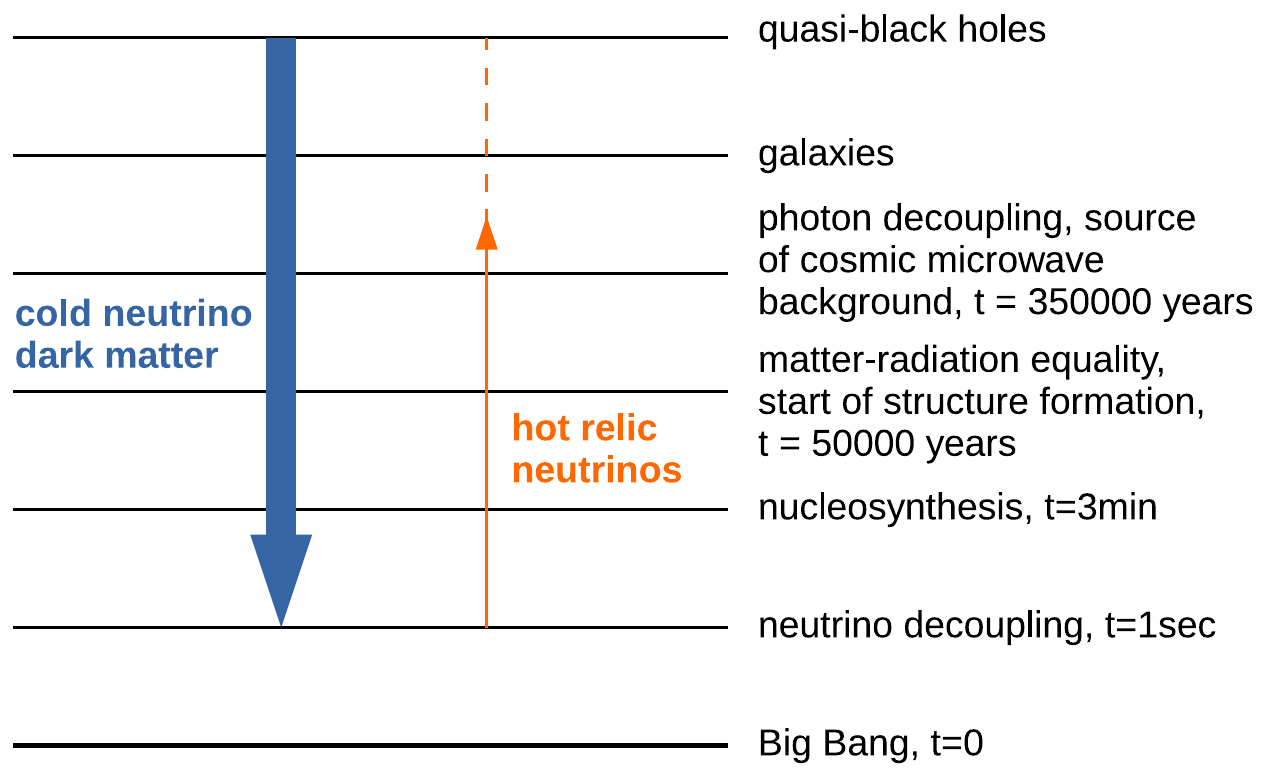}
\end{center}
\caption{An example scenario with cold neutrino dark matter emitted from quasi-black holes.}\label{f2}
\end{figure}

\section{Introduction}
A recent paper by the author \cite{icm2ps2020} examined a model in which dark matter (DM) was emitted from compact massive objects known as quasi-black holes (QBHs). This term denotes spherically symmetric solutions of the general theory of relativity with the inclusion of material terms of various types \cite{1612.04889,0902.0346,1401.6562,1409.1501}. Such objects look like a Schwarzschild black hole from the outside, but are structured differently inside. Typically, such solutions do not have an event horizon, but instead have a spherical region with a strong gravitational redshift. In the model \cite{icm2ps2020}, QBHs are stationary T-symmetric objects emitting DM into the future and into the past. This turns out to be possible due to the special structure of the light cones in the solution, see Fig.\ref{f1}. Because of this, near the Big Bang (BB), a significant amount of DM may come from QBHs located in the future, see Fig.\ref{f2}. This scenario will be discussed in detail in this work.

Time-backward DM was previously considered in \cite{cond-mat/9911101}. This matter was produced in Big Crunch (BC) and propagated in the direction of BB, obeying the reverse thermodynamic arrow of time. Today, such matter would consist of extinct stars and other cold objects that, by their gravity, could reproduce the effects of cold dark matter (CDM). Although the possibility of experimental detection of such objects by gravitational lensing was mentioned in \cite{cond-mat/9911101}, the connection with BB conditions was not analyzed. Presumably baryonic composition of such matter would lead to interaction, at least under the conditions of nucleosynthesis, which would greatly change its parameters.

This work will consider similar scenarios, but DM will be emitted from QBHs (with a possible BC contribution if present). In the main versions of the model, DM will be composed of cold neutrinos. Low-energy neutrinos interact very weakly with matter and pass through BB structures shown in Fig.\ref{f2}: the formation zone of cosmic microwave background (CMB), matter-radiation equality (MRE), Big Bang nucleosynthesis (BBN). The interaction begins only in the neutrino decoupling (ND) zone, at $t=1sec$.

The work \cite{1803.08930} considered a T-symmetric BB, in which the time-reverse part contained an alternative universe, CPT-conjugate to ours. A scenario was explored in which right-handed neutrinos were produced gravitationally in BB, forming a sterile neutrino CDM in both universes. Clearly, a different scenario is being considered here, although it should be noted that the light cone structure of T-symmetric BB and QBHs considered here are similar.

In this work, the model will be compared with the observed parameters, BB cosmology, neutrino experiments. Today, important observable features are galactic rotation curves (RCs, \cite{9502091,9503051,9506004,0703115,1609.06903,0811.0859,0811.0860,1110.4431,1307.8241}). These curves are usually described by empirical profiles \cite{NFW,Burkert,Einasto}, one can also use model solutions, which will be given below in Appendix~A. The question of the existence of systems with a backward arrow of time is disputed. On the other hand, there are many models of this kind \cite{wf1945,wf1949,vanStockum1937,Goedel1949,Kerr1963,Gott1991,MorrisThorne1988,1310.7985,gr-qc/0009013} and a general formalism for their description \cite{Visser1996}. Such models will be reviewed in Appendix~B.

Appendix~C considers the photonic version of the model, which can be closed when considering the interaction of EM-waves with interstellar/intergalactic medium (ISM/IGM). Based on the modeling \cite{BlandfordThorne,solar,KolobovEconomou,Weibel}, EM-waves of the required frequency are quickly suppressed by ISM/IGM, in addition, the pressure they exert would blow ISM out of the galaxy.

In the second section of the work, the cold neutrino case in the considered model with T-symmetric QBHs is studied in detail. The third section examines the possibility of solving general cosmological problems within the framework of the considered model. The main results are listed in the conclusion.

\section{Cold neutrino case reopened}

\paragraph{Contribution of QBHs.} The absence of an event horizon for QBH solutions is explained by the estimate $2GM/(rc^2)<1$, where $M$ is enclosed Misner-Sharp mass, $r$ is radius, $G$ is gravitational constant, $c$ is speed of light. For the Schwarzschild solution, $M$ is constant, and as $r$ decreases below the gravitational radius, the estimate is violated and an event horizon is formed. For QBHs, due to the matter equations, $M$ decreases with decreasing $r$, which corresponds to a positive mass density on the solutions. If $M$ decreases faster than $r$, then the estimate is satisfied and the event horizon is not formed. For solutions \cite{icm2ps2020}, $M$ decreases very quickly, due to the phenomenon of mass inflation for counterstreaming relativistic flows \cite{gr-qc/0411062}. Different models of QBHs have different structures; as general parameters, one can choose the initial energy of particles inside QBH and the gravitational redshift when particles exit outside. Internal energies can be very high, up to Planck values, but the redshift can move particles to ultra-low energies, making them virtually undetectable. Solutions for massless particles were considered in \cite{icm2ps2020}. For massive particles, the solutions are similar; they are characterized by the presence of turning points. If the energy, taking into account the redshift, turns out to be less than the rest energy of the particle, then the particle turns and remains inside QBH. If the energy is greater, then the particle flies out. If the initial energy has a distribution with a rapidly decreasing tail, then the output kinetic energy will have a distribution starting from zero and also having a rapidly decreasing tail. By choosing the parameters, it is possible to ensure that the output distribution is localized in the non-relativistic region.

In gravitationally coupled systems, the non-relativistic velocities of particles must be compared with the characteristic escape velocity. If particle velocities are greater than the escape velocity, particles can move between galaxies along almost straight radial trajectories. At the same time, the system considered in this model does not disperse in space, since it has T-symmetric incoming and outgoing flows that compensate each other. Such solutions are considered in Appendix~A as type A1. Radial flows for such solutions correspond to DM density, changing geometrically $\rho\sim r^{-2}$, which corresponds to a linearly growing mass $M\sim r$ and a constant orbital velocity $v=Const$, that is, flat RC.

When taking into account many QBHs, a linear superposition of such distributions can be used. Assuming that all black holes in the galaxy are QBHs, and also that the distribution of QBHs follows the distribution of luminous matter (LM), a highly concentrated distribution of QBHs towards the center will be obtained. Then, at large distances from the galactic center, the overall RC will be approximately flat. At small distances from the center, the distribution of DM will be modulated by the distribution of QBHs, and RC will deviate from the flat shape. The exact calculation was made in \cite{icm2ps2020}. The resulting model curves describe well the experimental data, both RCs of real galaxies averaged over a large set (Universal Rotation Curve, URC, \cite{9502091,9503051,9506004,0703115,1609.06903}), and for RC of the Milky Way over a wide range of distances (Grand Rotation Curve, GRC, \cite{0811.0859,0811.0860,1110.4431,1307.8241}).

\begin{figure}
\begin{center}
\includegraphics[width=0.7\textwidth]{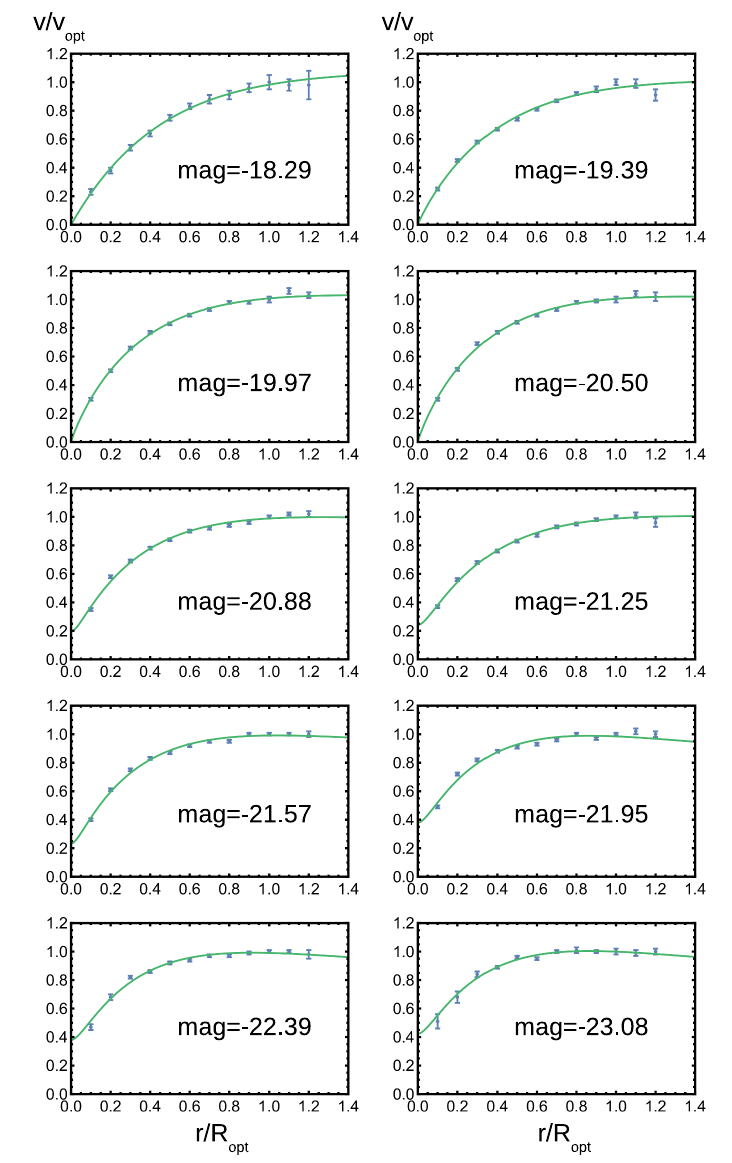}
\end{center}
\caption{Universal Rotation Curve (blue points with error bars), fitted by QBH-model (green curve). Reprinted from \cite{icm2ps2020} under CC BY 3.0.}\label{p1_f3}
\end{figure}

\begin{figure}
\begin{center}
\includegraphics[width=0.4\textwidth]{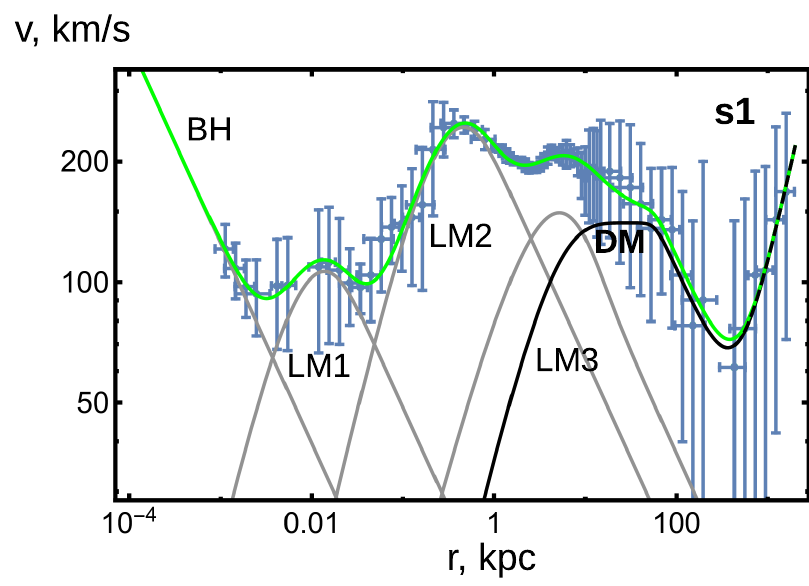}
\end{center}
\caption{Grand Rotation Curve (blue points with error bars), fitted by QBH-model (green curve). The black curve represents DM contribution, the gray curves are the contributions of different LM structures. Reprinted from \cite{icm2ps2020} under CC BY 3.0.}\label{p1_f2b}
\end{figure}

Fig.\ref{p1_f3} shows the model fit of \cite{icm2ps2020} experimental URC data from \cite{9502091}. According to the procedure \cite{9502091}, data on $\sim1000$ spiral galaxies are binned by the magnitude, normalized to the values $(r/R_{opt},v/v(R_{opt}))$, where $R_{opt} $ -- optical radius of the galaxy, containing 83\% of light, and averaged. The resulting experimental points are fitted with basis forms representing DM and LM contributions. DM was calculated using the QBH model \cite{icm2ps2020}, LM was modeled in the standard way \cite{freeman}. Fig.\ref{p1_f2b} shows the model fit of \cite{icm2ps2020} experimental GRC data from \cite{1307.8241}. The data represent the experimental Milky Way rotation curve. The fit included basis forms from \cite{icm2ps2020} for DM and several LM structures introduced in \cite{1307.8241}. Note that the fits in \cite{icm2ps2020} used relativistic DM particles, however, for non-relativistic particles in the regime of velocities greater than escape velocities, the same model curves will be obtained, so the results of the fit remain valid.

When considering the cosmological expansion of the universe, the relativistic model \cite{icm2ps2020} required a special cutoff at the halo boundary to ensure equivalence with the observed LambdaCDM regime. For the non-relativistic scenarios considered here, DM is cold, and, after adding an appropriate amount of dark energy (DE), automatically provides the observed expansion regime.

Let us also note that the intensity of emission of matter by QBHs into the past and future is in no way related to the mass of the source; in particular, the linearly growing mass of the halo $M\sim r$ can significantly exceed the mass of QBHs emitting it. On the other hand, at small distances the main contribution to gravitational acceleration comes from the masses of sources that determine the orbital motion of nearby bodies around QBHs. The contribution of DM to gravitational acceleration, after summing over all sources, appears only on galactic scales.

Another solution, considered in Appendix~A as type A2, corresponds to the case when the trajectories of DM particles are curved and, being emitted into the future or into the past, no longer return to the sources. Instead of radially directed flows, chaotic motion arises, which can be described using the well-known model of isothermal halo. Solutions of both types are stationary, and the energy flux through spheres of radius $r$ vanishes, which corresponds to the conservation of mass for each $r$. For solutions A1, as long as QBHs exist, half of the particles are emitted into the future, half into the past, forming incoming and outgoing radial flows of equal intensity. For solutions A2, particles emitted into the future gradually replace particles emitted into the past. As time moves forward, there are more and more particles emitted into the future, and fewer and fewer particles emitted into the past. For both types of solutions, only particles emitted from QBHs into the past are present near the BB. So far, the consideration has not depended on the type of particles; further on in this section, neutrinos will be considered as DM particles.

\paragraph{Neutrino mass.} When considering light neutrinos, the neutrino mass $m_\nu=0.1eV$ will be used in the estimates. It will be assumed that all 6 neutrino species (3 families $\times$ 2 particle/antiparticle) have the same masses, also in the considered scenarios all species have the same density. This mass value fits into the lower limit for neutrino oscillations, from which it follows that there must be at least one eigenstate with a mass of $m_\nu>0.05eV$. The value used falls within the upper limit of the KATRIN experiment \cite{KATRIN2022} $m_\nu<0.8eV$. This value does not fit into a recent analysis \cite{2106.15267} of Planck satellite data in combination with other experiments, which led to the strongest constraint to date $m_{\nu,sum}<0.09eV$. The neutrino mass estimate used in this work can be lowered to satisfy this constraint as well, yielding qualitatively similar results. However, this restriction can no longer be satisfied for neutrino masses equal in 3 families. In this work, the feasibility of neutrino scenarios with the simplifying assumption that neutrino masses are equal will be evaluated and $m_\nu=0.1eV$ will be fixed.

\paragraph{Matching with Big Bang cosmology.} In the standard LambdaCDM model, the case of light neutrinos as DM candidates was excluded. The main argument is the calculation of the density of relic neutrinos in homogeneous state, which gives $n_{\nu,uni,std}=56\cdot10^6m^{-3}$ for each of the 6 neutrino species. For $m_\nu=0.1eV$ the total mass density will be $\rho_{\nu,uni,std}=6\cdot10^{-29}kg/m^3$, which is significantly less than the estimated uniform DM density in the universe $ \rho_{dm,uni}=2.7\cdot10^{-27}kg/m^3$. At the same time, in LambdaCDM the neutrino density cannot be freely increased. This density is related to other measured densities by conditions of thermodynamic equilibrium. Increasing the density of hot neutrinos emitted from BB will change the overall radiation density in early cosmological stages. This density affects BBN outcome, CMB characteristics, and can push MRE to later stages, leaving little time for structures to form. Because of this, DM scenario with cold neutrinos was abandoned in LambdaCDM.

The model considered here has an alternative source of cold neutrinos -- QBHs. It has unusual location in the future; this question will be discussed in detail in Appendix~B. If DM consisted of cold classical particles ($T=0$, dust), then when they propagated from the future to the past, in the direction of BB, DM would remain cold. Like other types of CDM, for most of its existence, cold neutrinos interact with other types of matter only gravitationally. Light neutrinos begin to interact actively with other matter in ND zone, $t\sim1sec$ after BB.

Let us consider this process from the point of view of an observer related with the neutrino flow, with the arrow of time directed into the past. A large number of cold neutrinos pours into the equilibrium hot plasma of neutrinos, photons, electrons and positrons. Then, the temperature and the number of neutrinos are equalized, coming into the equilibrium with other components of the plasma. Now let us consider this process in the forward direction of time that is conventional to us. In the initially equilibrium plasma, the formation of a large number of cold neutrinos begins, which are separated and then exist in the form of CDM. This process proceeds contrary to thermodynamic principles, since it obeys thermodynamics with the backward arrow of time. Appendix B discusses this question in detail. It is sufficient that the model is logically consistent for at least one direction of time. From the point of view of an observer with the forward direction of time, the formation of cold neutrinos necessary to describe DM occurs, not constrained by the conditions of thermodynamic equilibrium.

At ND moment, hot neutrinos are also separated, being in thermodynamic equilibrium with the plasma; the number of hot neutrinos corresponds to the usual calculation. These hot neutrinos evolve exactly as in the standard model. Due to the action of the cosmological redshift, at some moment, depending on the neutrino mass, hot neutrinos stop and become cold. This portion of cold neutrinos then merges with others, representing a small addition of relic neutrinos to the CDM.

To summarize, in the considered model, CDM consists of cold neutrinos emitted from QBHs, not bound by thermodynamic equilibrium conditions with BB. This CDM spends all stages from today to ND in a cold state, interacting with other types of matter only gravitationally. All LambdaCDM relations in this time interval are satisfied. Only events before ND are subject to modification, where CDM dissolves in the equilibrium plasma and disappears.

\paragraph{Quantum restrictions.} Especially for light neutrinos, quantum restrictions must be taken into account. These restrictions follow from Pauli exclusion principle (PEP) and are the second, stronger argument why neutrinos were excluded as possible candidates for the role of DM particles. To explain the observed $\rho_{dm,uni}$, assuming $m_\nu=0.1eV$, the equality of masses and densities for neutrino species, the density of one species $n_{\nu,uni}=2.5\cdot10^9m^{-3}$ is required. According to PEP, this density in the non-relativistic case cannot exceed the value $n_{PEP}=(m v_F)^3/(6\pi^2\hbar^3)$, where $v_F$ is the maximum speed corresponding to Fermi energy $\epsilon_F=mv_F^2/2$. Assuming $v_F=100-1000km/s$ in the range of escape velocity in galaxies, $n_{\nu,PEP}=8\cdot10^4-8\cdot10^7m^{-3}$, the uniform density required to describe DM turns out to be above this limit.

PEP restrictions from the homogeneous case are quite soft. When considering DM, galaxies have an increased density, $\rho_{dm,loc}=7.7\cdot10^{-22}kg/m^3$ from \cite{1003.3101}, which is also in agreement with recent estimations \cite{2012.11477}, which corresponds to $n_{\nu,loc}=7.2\cdot10^{14}m^{-3}$. Thus, $n_{\nu,PEP}$ turns out to be exceeded by more than 7 orders of magnitude. Also, when considering the compression of matter during backward evolution in BB direction, PEP can only be satisfied by increasing $v_F$ to relativistic values. This will lead to significantly increased radiation density in the early stages, which will greatly change the observed evolution of the universe.

To implement the considered scenario, the only option seems to be to make PEP no longer applicable. The most radical proposal is that neutrinos, although they have spin 1/2, obey Bose statistics. As analyzed in detail in \cite{hep-ph/0501066}, some aspects of the standard model will have to be changed, and it is unknown whether these modifications can describe all the observed data. In this work, another option will be considered. As is known, the presence of an arbitrarily small attractive interaction between fermions in PEP regime leads to the formation of Cooper pairs. Such interactions were considered in \cite{0911.5012,1008.5214,2004.03731}. Following this suggestion, let us assume that neutrinos can form Cooper pairs and undergo Bose-Einstein condensation (BEC).

Possible solutions of this type are considered in Appendix~A. Solutions of type A1 repeat the classical solutions with radial flows, with a semi-classical representation for the wave function of the coherent state. In this case, the flows represent global motion, and local transverse and longitudinal thermal disturbances are frozen ($T=0$). Solutions of type A2 are boson stars described in \cite{Chavanis1}. Typically, such solutions use a different mass region $m\sim10^{-22}eV$, more suitable for axion-like particles (ALPs). However, in the model \cite{Chavanis1} it is possible to configure the mass to the required range if one enables the repulsive interaction between the pairs. The equilibrium state has the form of a degenerate BEC core and an isothermal halo surrounding it. In addition, slowly relaxing non-equilibrium states are possible. The structure of solutions A1-2 describes flat galactic RCs, which to a first approximation coincide with observations. The models also have the ability to perturb curves, obtaining deviations from the flat case. The models should be extended by LM contributions as well as the distribution of QBHs across the galaxy, similar to \cite{icm2ps2020}. Being continued into the past until the formation of structures, the solution becomes homogeneous and forms an equilibrium BEC or non-equilibrium cold Bose gas. It remains the same when the density increases to ND, before which the neutrino DM disintegrates and dissolves in the equilibrium BB plasma.

Another class of solutions is formed by neutrinos from extensions of the standard model. Such neutrinos have (almost) zero direct interaction with ordinary matter and are designated as sterile. Usually they are assigned larger mass values. The work of \cite{Chavanis2} considered particles of this type in the mass range $165eV-386keV$. In further estimates, $m_{s\nu}=100keV$ will be fixed. This value allows to bypass PEP limitation $n_{s\nu,PEP}=8\cdot10^{25}m^{-3}$, as for the homogeneous estimate $n_{s\nu,uni}=7.6\cdot10^3m ^{-3}$, and for the local estimate $n_{s\nu,loc}=2.2\cdot10^9m^{-3}$. Here the presence of two species of sterile neutrinos is assumed (1 family $\times$ 2 particle-antiparticle) and $v_F=1000km/s$ is selected. Being continued into the past $n_{s\nu}=n_{s\nu,uni}(1+z)^3$, to BBN $z_{BBN}=4\cdot10^8$, $n_{s\nu ,BBN}=4.8\cdot10^{29}m^{-3}$, PEP with a high value of $v_F=10^8m/s$ does not apply yet, $n_{s\nu,PEP}(v_F)=8\cdot10^{31}m^{-3}$. At earlier stages, sterile neutrinos of the considered mass should form a degenerate relativistic Fermi gas.

As a result, throughout the evolution from today to BBN, sterile neutrinos with a chosen mass can remain non-relativistic, and the classical RC profiles A1-2 can be used to describe the galaxies today. The work of \cite{Chavanis2} also considers quantum fermionic stars. They also consist of a degenerate core and an isothermal atmosphere. PEP operates at higher densities $n_{s\nu,loc}$ achievable in the vicinity of the core. Note also that many models consider much heavier sterile neutrinos, for example $m_{s\nu}=4.8\cdot10^8GeV$ in \cite{1803.08930}. For such values of masses, PEP restrictions are shifted to even earlier stages of BB, and disappear in galaxies. In practice, very heavy neutrinos behave like classical non-relativistic particles, for which the classical solutions of A1-2 type can be used in the considered model.

\paragraph{Matching with neutrino experiments.} KATRIN experiment \cite{2202.04587} not only provides an upper limit on the neutrino mass, but also an upper estimate of the neutrino background density. Usually the background is expressed in terms of overdensity for one neutrino species $\eta=n_\nu/n_{\nu,uni,std}$. In this expression, the corresponding local density DM for $m_\nu=0.1eV$ is $\eta=1.3\cdot10^7$. KATRIN experimental provides a limit $\eta<9.7\cdot10^{10}/\alpha$ and planned maximum sensitivity $\eta<10^{10}/\alpha$ at 90\% CL, where $\alpha=1$ for Majorana neutrinos, $\alpha=0.5$ for Dirac neutrinos. The upper limits are significantly higher than the estimated signal for neutrino CDM.

The planned IceCube-Gen2 experiment \cite{2207.02860} can measure the annihilation reactions of cosmic high-energy neutrinos against the cosmological neutrino background. In this case, values of $\eta\sim10^{11}$ can be achieved for redshifts $z=2$, which is comparable to KATRIN limits.

The planned experiment PTOLEMY \cite{2111.14870} is aimed at significantly increased sensitivity to the neutrino background level. The estimated sensitivity $\eta\sim1$ at $m_\nu=0.1eV$ for a 1-year exposure should be sufficient to detect a signal at $\eta=1.3\cdot10^7$ level in the considered cosmological scenario.

\paragraph{Possible reduction of the signal.} According to estimates \cite{2111.14870}, the sensitivity of PTOLEMY experiment deteriorates significantly as the mass decreases to the region $m_\nu<0.1eV$. Let us also recall that KATRIN and PTOLEMY experiments use the reaction of electron neutrino capture by a tritium nucleus. The estimates above were made under the assumption of equality of masses and densities for neutrino species. For large mass values, the difference in mass between species is limited by oscillation experiments and is small. At low masses, stronger differences are possible. In addition, a violation of the equality of densities, up to the complete absence of a signal for electron neutrinos, is theoretically possible. Although such asymmetry will require special explanations.

The more important factor is the following. In the considered model, some neutrinos are emitted from QBHs forward in time ($\nu_f$), and some are emitted backward in time ($\nu_b$). There is also a small relic neutrino contribution from BB ($\nu_{BB}$) and a possible unknown contribution from the future BC ($\nu_{BC}$), see discussion below. Capture or annihilation experiments can only detect neutrinos that have sources in the past. Indeed, capture or annihilation destroys the neutrino and makes it impossible for it to propagate to future sources. For solutions of type A1, the contributions $\nu_f$ and $\nu_b$ from QBHs are equal in the time interval while QBHs exist. If one ignores the contributions of $\nu_{BB}$ and $\nu_{BC}$, this will lead to a 2-fold reduction of the signal. For solutions of type A2, the contribution $\nu_f$ increases with time, while the contribution $\nu_b$ decreases. This dependence can be modeled as $n(\nu_f)/(n(\nu_f)+n(\nu_b))=t/t_m$, where $t_m$ is the minimum of the lifetime of the universe and the lifetime of QBHs. For $t$ corresponding to today and large $t_m$, the signal reduction factor can be arbitrarily small. In this scenario, when the universe is relatively young, almost all DM is produced in its future. Adding a significant contribution from $\nu_{BC}$ can further reduce this estimate.

When considering sterile neutrinos, similar estimates apply. It is assumed that neutrinos are not completely sterile, but still have a small direct interaction with ordinary matter, enough to be detected. There is a factor in the model that can significantly reduce the signal and prevent such registration in absorption and annihilation experiments. Registration is possible in T-symmetric channels, for example, scattering. Otherwise, DM manifests itself only in gravitational effects.

\begin{figure}
\begin{center}
\includegraphics[width=0.7\textwidth]{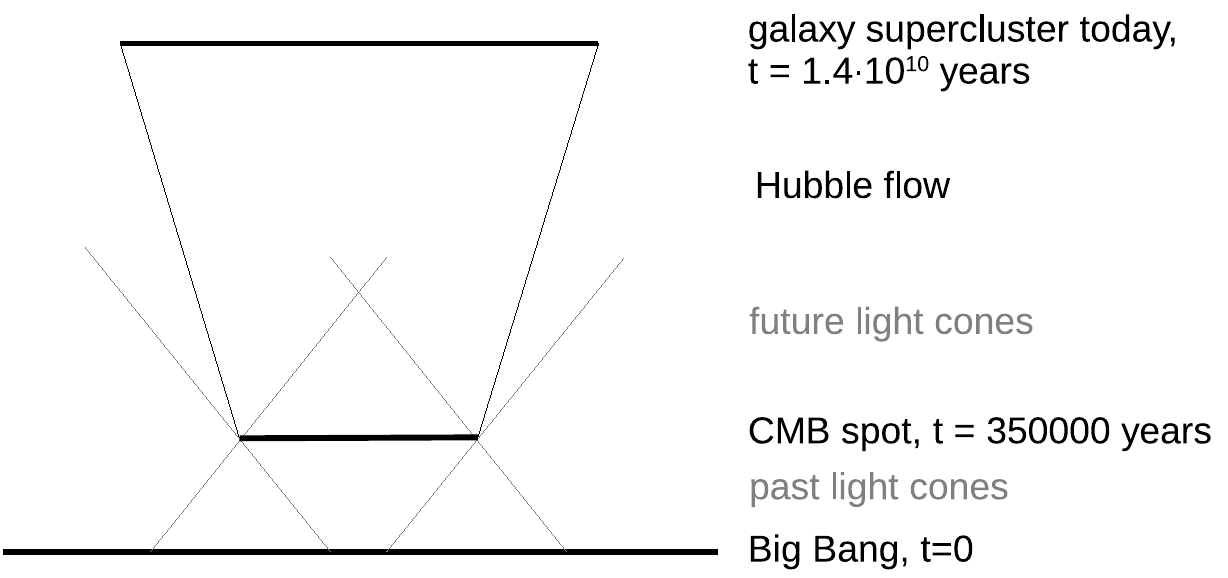}
\end{center}
\caption{To consideration of horizon problem.}\label{f6}
\end{figure}

\section{Relation to cosmological problems}
The direction of DM evolution from the future to the past as perceived near BB, typical for the considered model, leads to a number of consequences related to known cosmological problems. In this section, the consideration is general, independent of the type of DM particles.

\paragraph{Horizon problem.} This problem arises in the standard cosmological model and consists in the fact that the experimentally observed correlated regions in CMB are too large, see Fig.\ref{f6}. Their continuations along the light cones of the past in the direction of BB correspond to causally disconnected regions. The solution of the problem in the considered model is that DM sources are in the distant future, and continuations along the future light cones have non-zero intersections. In fact, it is not even necessary to consider the light cones. In backward evolution, galactic superclusters are transported by Hubble flow and leave imprints in CMB, in the form of spots of the size that is observed now. With further backward evolution towards BB, correlated points in CMB are not causally related to common events at the moment of BB, but they are causally related to common sources in the distant future, that is why they are correlated.

\paragraph{Flatness problem.} For simplicity, in further discussion, DE will be turned off in the distant future of the universe. Specific mechanisms for such shutdown will be discussed below. Let us consider DE as a transient phenomenon that disappears over time or is transformed into DM. In this case, the evolution of the universe will be mainly determined by DM. Three cases are known, a closed universe with BC $\rho>\rho_{crit}$, an open universe $\rho<\rho_{crit}$ and a borderline flat universe $\rho=\rho_{crit}$. In the considered model, for a closed universe, if its lifetime is short, QBHs will produce too little DM for the universe to collapse. For an open universe, QBHs will produce too much DM, infinitely much, or limited by the lifetime of QBHs. It can be seen that the model has a feedback that can stabilize the solution on the boundary case. The formal expression of this feedback is considered in Appendix B. For the closed case, the lifetime of the universe turns out to be inversely proportional to the intensity of QBHs emission, and at low intensity the solution tends to the flat case.

Note that both problems, horizon and flatness, are solved in the standard model using the inflation mechanism. The considered model can also include an inflationary period in the early stages of BB, although its presence is not necessary to resolve the problems described above.

\paragraph{Cosmological constant problem.} This is the most complex problem, involving several questions \cite{Blau2022}. First, why the cosmological constant $\Lambda$ or DE density determined by it cannot be significantly greater than its actual value. A detailed analysis shows that in this case galaxies would not have formed \cite{Tong2019}. Further, a reference to the anthropic principle is possible: life as we know it requires stars and the galaxies forming them to exist. If there were no life, there would be no one to raise the question about the value of the cosmological constant. Note that in the considered model, the upper limit of the cosmological constant is not actually related to the anthropic principle. If, at a fixed DM density, the cosmological constant were greatly increased, then there would be no galaxies and stars, as well as black holes and DM they produce. Thus, the value of the cosmological constant is limited from above not by the anthropic principle, but by the self-consistency of the model.

The second question is why the cosmological constant does not take smaller values, in particular, why it cannot be put to zero. Within the framework of the considered model, let us try to answer this question as follows. Consider backward evolution, in which galaxies are initially at large distances, then come closer and are pressed into each other. At the moment when galaxies begin to touch the outer parts of their halos, forces can begin to act between them due to the redistribution of DM flows in them. These forces can act like friction, slowing down the approaching of galaxies. In forward evolution, such slowdown is equivalent to the accelerated recession of galaxies. Thus, the friction force in backward evolution can emulate the effect of the cosmological constant in forward evolution. In answer to the question posed above, the value of the cosmological constant today cannot be reduced or turned off, since in the considered model it is not a free parameter; its value is related with the effects of DM redistribution in galaxies as they approach each other in backward evolution. Such a mechanism could explain the nature of DE and would be a direct signal of the backward time arrow in the evolution of DM. It would be interesting to perform a quantitative evaluation for such a cosmological scenario.

Note that identifying the cosmological constant with the forces that arise when galaxies approach each other also explains why the cosmological constant can be turned off in the distant future of the universe, as required above in the analysis of the flatness problem. There are other options for such shutdown. When considering a scalar field to explain the cosmological constant, in particular, in the quintessence model, the following possibility arises. In the distant future, a phase transition may occur, the gravitational condensation of the scalar field from the homogeneous state describing DE, with the formation of isolated boson stars of very large size. Such stars, surrounded by vacuum, will behave like CDM on a large scale. Thus, effective transition of DE into DM can occur. Another possibility within the quintessence model is a self-generated shutdown of the acceleration of the universe at large time, similar to the shutdown of the inflation field in the early stages of the evolution of the universe. Possible transition processes between DE and DM, as well as direct modification of the DE contribution, were considered in a number of works \cite{1808.02472,2012.01407,1809.05678,1812.06854,1911.04520,1903.02370,2010.10823,2002.06127,1804.08558,1812.03540,1907.12551,2001.05103,1908.04281,1910.09853,2009.12620,1902.10636,1908.09843,2002.03408,1908.03324,1907.01496,1906.09189,1701.08165,1807.03772}.

There is also a third question related to the cosmological constant, known as the coincidence problem. It is necessary to explain why the mass densities of DM and DE, which have different power-law behavior on the scale factor of the universe, are equal in order of magnitude today. It is difficult to answer this question without resorting to anthropic considerations, since ``today'' represents here the current era of development of our civilization. More precisely, the moment of matter-Lambda equality (MLE) occurs slightly earlier $t_{MLE}=10^{10}$~years than today $t_0=1.4\cdot10^{10}$~years. This approximate coincidence could have happened by chance. It can be more indirectly described as a random covering of MLE moment by an interval $(0.4-2)\cdot10^{10}$~years of the preferential existence of yellow dwarfs (G-type main sequence stars), to which our Sun belongs. Thus, all inhabitants of such systems can predominantly register the coincidence of MLE with their today moment, in terms of mass density with an accuracy of $\pm1$dex. It can also be assumed that intelligent life exists on planets near red dwarfs (M/K-type main sequence stars), the era of which lasts trillions of years. Their inhabitants will most likely observe MLE in the distant past from their today moment, just as for us MRE is located in the distant past.

\paragraph{Hubble tension.} There is experimental indication that in the early epochs of the evolution of the universe there were additional relativistic degrees of freedom, which then disappeared. To describe this possibility, consider the following modification of the model. According to the influence of cosmological redshift presented in Appendix~B, DM is not necessarily cold throughout the evolution of the universe. In the early moments, DM may have been hot and had an energy that was not in thermodynamic equilibrium with BB plasma. This energy can be used as a free parameter, or the distribution of this energy as a free functional degree of freedom. In the process of further evolution, DM cooled and became cold at a moment depending on the initial energy and mass of the particles. As a result, at the early stages of evolution it is possible to modify the radiative density, and at the moment today DM can be left cold, non-relativistic. New degrees of freedom that appear can be included in the fit of the extended model with respect to the observed data. This may affect existing tensions in the standard model, such as Hubble tension.

Let us illustrate this possibility with a concrete example. When considering the backward evolution in the direction of BB, the galaxies come closer and are pressed into each other. At this moment, part of DM can flow outside the galaxies, forming a cold free streaming DM component. Assuming the speed of DM particles after escaping the galaxies in the range $v\sim100-300km/s$ today, applying the cosmological redshift described by the formulas from Appendix~B, obtain relativistic speed values at $z\sim1000-3000$, in the range between CMB and MRE. Before this epoch, there are additional relativistic degrees of freedom in the system, which can affect the Hubble tension.

\section{Conclusion}

The paper considers two dark matter scenarios with low-energy neutrinos emitted from T-symmetric quasi-black holes. In the first, light neutrinos pair to bosonic quasiparticles that form massive halos around galaxies. In the second, heavier sterile neutrinos form massive halos without pairing. At the observable level, both scenarios appear to be equivalent to the standard LambdaCDM model. In the considered model, there are additional sources -- quasi-black holes -- that can produce the required number of neutrinos in addition to the Big Bang. This allows to bypass the restrictions on the number density of light relic neutrinos present in the standard model. The expansion of the universe registered today occurs in the same way as for cold dark matter of any kind, mixed with an appropriate amount of dark energy. The considered model solutions for dark matter halos produce galaxy rotation curves similar to the basis shapes used previously in successful fits of the observed data. This ensures that the model rotation curves match the observations.

The paper discusses the matching of the model with direct detection neutrino experiments. The neutrino density required to describe dark matter satisfies constraints from the ongoing KATRIN experiment and the planned sensitivity of the IceCube-Gen2 experiment. The planned PTOLEMY experiment has a sensitivity sufficient to detect the neutrino background described in the model. On the other hand, the model contains a mechanism that can reduce the signal in neutrino absorption or annihilation experiments, leaving the gravitational influence of dark matter unchanged.

As an additional option, DM consisting of low-energy photons is considered. This option can be closed when considering the interaction of EM-waves with the interstellar/intergalactic medium, which absorbs low-frequency photons. In addition, the pressure exerted by the photons would blow the interstellar medium out of the galaxy.

The work also discusses the relation of the model to general cosmological problems. Placing dark matter sources into the distant future provides a possible solution to the horizon problem. The model provides a mechanism for self-regulation of the dark matter mass and the lifetime of the universe, directly related to the flatness problem. This self-regulation mechanism, as well as the possible interaction between galaxies during their approach and the position of the star formation epoch, are responsible for different aspects of the cosmological constant problem. Modifications of the model at early cosmological stages with the addition of new tuning parameters can remove the cosmological tensions present in the standard model.

\footnotesize

\appendix

\section{Model solutions for dark matter halo}
\paragraph{A1: Radially emitted dark matter.} Scenarios with radial DM flows in the relativistic case were considered in \cite{icm2ps2020}. In this work, the non-relativistic case will be considered. A stationary spherically symmetric self-gravitating solution with radial DM flows is described by equations of the form
\begin{equation}
dM/dr=8\pi c_1/u,\ du/dr=-GM/(ur^2),
\end{equation}
where $r$ is the radius, $M$ is enclosed mass, $u$ is the speed of the radial flow, $G$ is the gravitational constant, $c_1>0$ is an arbitrary constant that specifies the intensity of the flow. This system can be obtained from 3D Euler equations for spherically symmetric flows. Physically, the mass function corresponds to the density $\rho_1=c_1/(ur^2)$ specified for one flow (for example, outgoing), the same density describes another (incoming) flow, hence the double factor in the first equation. The second equation represents the energy condition for the trajectory of an individual DM particle: $\varphi+u^2/2=Const$, where the gravitational potential is $\varphi=\int dr GM/r^2$.

By changing the variables $r=\exp x$, $GM/r=y$, $c_2=8\pi Gc_1$ the equations are reduced to the form
\begin{equation}
dy/dx+y=c_2/u,\ du/dx=-y/u.
\end{equation}
The equations are autonomous, invariant under shifts $x\to x+Const$, which corresponds to scaling $r\to r\cdot Const$ for the argument of the original system. By dividing the equations by each other, one can obtain a single differential equation $dy/du=u-1/y$, whose general solution can be written in terms of Airy functions. However, direct numerical integration of this system is a simpler way to describe the solutions. Another symmetry of the system is the scaling of variables $u\to ua$, $y\to ya^2$, $c_2\to c_2a^3$.

Convenient parameterization of solutions can be carried out as follows. Due to the autonomy of the system in $x$-coordinates, the initial value $x_0=\log r_0$ can be set arbitrarily, in accordance with the initial radius $r_0$, which sets the scale for the geometric dimensions of the system. The orbital speed is determined by the relation $v=y^{1/2}$. It is possible to fix the condition $y'(x_0)=0$, while $x_0$ will determine the position of the minimum orbital velocity $v_0=y_0^{1/2}$. The remaining free parameter is the initial velocity of DM particles $u_0$, through which the constant $c_2=u_0y_0$ is also determined. The solution can be carried out in {\it Mathematica} system using {\tt NDSolve} method, with initial data $y(x_0)=y_0$, $u(x_0)=u_0$, until the threshold $u=0$ or the upper limit $r=r_1$ is reached, depending on what happens first. Next, the values $r_0=10kpc$, $v_0=150km/s$, $r_1=250kpc$ are used.

\begin{figure}
\begin{center}
\includegraphics[width=0.7\textwidth]{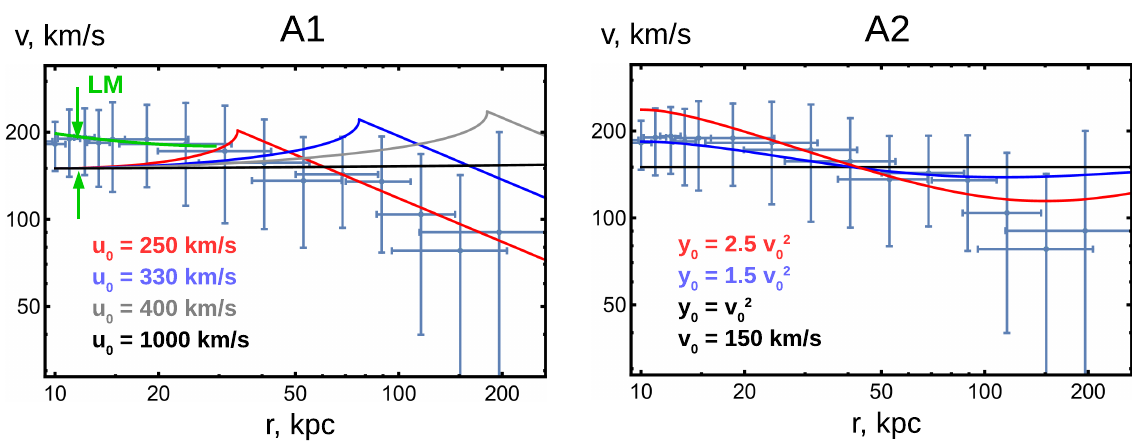}
\end{center}
\caption{External part of GRC: model solutions superimposed on observational data.}\label{f3}
\end{figure}

In Fig.\ref{f3} on the left, typical solutions of this system of equations are superimposed on GRC observational data \cite{1307.8241} for Milky Way galaxy. In this work, a detailed fitting of the curves to the experimental data is not carried out, only a visual comparison. The outer part of RC $r\in[10,250]kpc$ with the predominant contribution of DM-halo is considered, since QBH modeling with a central source can be used for this range. In this modeling, the geometric distribution of QBHs follows LM distribution, which is concentrated at the center of the galaxy, so all QBHs can be effectively transferred to the center of the system. In Fig.\ref{f3} on the left, the slight rise of RC above DM curve at $r\sim10kpc$ is due to the contribution of LM, cf. Fig.\ref{p1_f2b}. Structures at $r>250kpc$ correspond to the influence of the neighboring galaxy M31/Andromeda and are cut off in this analysis, since isolated galaxies are considered in the modeling used.

The $v(r)$ graph from the considered model has characteristic peaks with a vertical derivative when the threshold is reached, which are changed to Newtonian drop $v^2=GM/r$ for a constant mass $M$. The parameter $u_0$, which determines the initial velocity of DM particles, affects the height and position of the peak. Two solution modes can be distinguished. At $u_0\sim200km/s$ the graph approximately fits into the error corridor, which is very large for the outer RC part. However, no sharp peak is visible in the observations. This may be due to large errors and the large width of the bins along the radius, which can blur thin structures. As $u_0$ increases to $300-400km/s$, the agreement between the graphs worsens. At high values of $u_0\geq1000km/s$, the peak moves out of the considered range, and the graph becomes approximately flat. This is a well-known idealized result, which also appears in the version with an isothermal halo considered below. The flat curve also approximately fits into the error corridor, only slightly exceeding the $1\sigma$ value at one point.

When considering solutions with a threshold pushed to large distances, it should be kept in mind that other effects begin to operate there. These include the influence of neighboring galaxies, DE contribution, and the cosmological expansion of the universe (Hubble flow). These effects can completely erase the mass peak and other threshold features of the considered solutions.

{\it Quantum solutions} are described by the spherically symmetric Schrödinger equation
\begin{equation}
i\hbar\df\psi/\df t=-\hbar^2/(2m)(r^{-2}\df/\df r)(r^2\df/\df r)\psi+m\varphi\psi
\end{equation}
with gravitational potential $\varphi=\int dr GM/r^2$, mass function $M=\int dr 4\pi r^2\rho$ and mass density $\rho=m|\psi|^2$. At $r=r_0$, a boundary condition on the momentum $p_m=mu$ is imposed, with the speed $u$ taken from the classical solution. Such momentum values correspond to de Broglie wavelength $\lambda=2\pi\hbar/p_m$, for the mass of the Cooper pair $m=0.2eV$ and $u=100-1000km/s$ obtain $\lambda=2-20mm$. Since this value is much smaller than the galactic size, the semiclassical approximation is applicable in this problem. The wave function of the coherent state \cite{Chavanis1} has the form $\psi=n^{1/2}\exp(iS(t,r)/\hbar)$, where $S$ is the action on classical trajectories. This formula describes the Madelung transformation $n=|\psi|^2$, $u=\df S/\df r/m$, which is exact, reducing Schrödinger equation to Euler equations. In the case under consideration, the quantum potential can be neglected due to Heisenberg uncertainty principle $Q=O(\hbar^2)$. Elementary calculations lead to $\phi=S/\hbar=-Et/\hbar+\int dr p_m/\hbar$. Here the first term represents the dependence of the phase on time, typical for stationary states with energy $E$. The second term represents the rapidly changing phase in the radial direction, corresponding to a spherical wave with momentum $p_m$. For the semiclassical approximation, this phase change dominates the calculation of the radial derivatives, reducing Schrödinger equation to the classical relation $E=p_m^2/(2m)+m\varphi$. The described solution represents global radial motion, while local, including transverse, motions are frozen (BEC with $T=0$). In this case, one needs to consider the superposition of incoming and outgoing waves. This leads to the wave function $\psi\sim n^{1/2}\exp(-iEt)\cos\phi$, which is a solution of Schrödinger equation with the potential averaged over oscillations. RC shapes obtained for quantum solutions coincide with the classical ones.

\paragraph{A2: Isothermal halo.} In this model, a stationary spherically symmetric solution is described by the equation
\begin{equation}
(-2 r w + G M) dM/dr + w r^2 d^2M/dr^2=0,
\end{equation}
also following from Euler equations in the absence of global flows and the presence of pressure $p$ associated with the mass density $\rho$ by an equation of state of the form $p=w\rho$. The constant $w=k_BT/m=<u_T^2/3>$ is expressed in terms of temperature $T$, particle mass $m$, Boltzmann constant $k_B$, thermal particle velocity $u_T$. Euler's equations are reduced to the condition of hydrostatic equilibrium $dp/dr=\rho g$, where $g=-GM/r^2$ is gravitational acceleration, and together with the condition $dM/dr=4\pi r^2\rho$ give the above equation for the mass function.

By changing the variables $r=\exp x$, $GM/r=y$ the equation is reduced to the form
\begin{equation}
(-2 w + y)y+(-w + y)dy/dx+w\,d^2y/dx^2=0.
\end{equation}
The equation is autonomous and has the symmetries $x\to x+Const$, $r\to r\cdot Const$; $y\to ya$, $w\to wa$. As in the previously considered model, the orbital speed is determined by the relation $v=y^{1/2}$. The equation has a particular analytical solution $y=2w$, which corresponds to the linear mass function $M=2wr/G$, inverse square density $\rho=w/(2\pi Gr^2)$ and constant orbital velocity $v=(2w)^{1/2}$. This is a well-known solution describing flat RCs, and is often considered as a first approximation for modeling of real RCs.

Less known is the fact that this answer is only a partial solution of the isothermal model. The above equations are of the second order and their solutions have two degrees of freedom, while in this particular solution for a given $w$ there are no additional degrees of freedom. One of the degrees of freedom is associated with the choice of the initial value $y(x_0)=y_0$, different from $2w$. Another may be related to the choice of the derivative $y'(x_0)$. Alternatively, as in the previously discussed model, one can fix the condition $y'(x_0)=0$ and use the degree of freedom $x_0$ to shift the argument. Numerical integration of solutions was carried out for $r_0=10kpc$, $v_0=150km/s$, $w=v_0^2/2$ and different values of $y_0$. The result is shown in Fig.\ref{f3} on the right.

The choice $y_0=v_0^2$ corresponds to the flat RC. Two other values $y_0=1.5v_0^2$ and $y_0=2.5v_0^2$ correspond to an increase in the initial mass for the radius $r_0$. This increase, in particular, can be caused by LM contribution, which dominates at smaller radius values. It is noteworthy that the deepening at values of $r\sim150kpc$, which in other types of modeling requires the introduction of special mechanisms, in this model is a consequence of oscillations near the flat RC, which are produced by the equations themselves.

{\it Quantum solutions} were presented in the works \cite{Chavanis1,Chavanis2}, the so-called boson and fermion stars. In quantum solution A1, gravitational forces were balanced by gas pressure flowing out of QBHs. In the quantum solution A2, they are balanced by the pressure from Heisenberg principle for boson stars and PEP pressure for fermion stars. Both types of solutions can be configured to the same resulting structure, a core of degenerate gas surrounded by an isothermal halo. For boson stars \cite{Chavanis1}, for particle masses in the considered range, a repulsive interaction between Cooper pairs is required. In this case, gravity is balanced by pressure due to repulsive forces, and the result depends not on the mass of the particles, but on the interaction constant at the repulsive quartic potential. For fermion stars in \cite{Chavanis2}, solutions are available in the range $165eV-386keV$, which is suitable for the sterile-type neutrinos considered here.

{\it Non-equilibrium states.} Note also that for light paired neutrinos in the configuration described above, the isothermal halo may have insufficient density to describe DM, due to the same quantum restrictions. Theoretically, these restrictions can be circumvented as follows. The formation of boson and fermion stars occurs due to the hypothetical mechanism of violent relaxation \cite{Chavanis1,Chavanis2}. If this mechanism does not operate during the time intervals under consideration, non-equilibrium bosonic solutions of arbitrary density are possible, in particular, literally repeating classical solutions of type A2. Non-equilibrium fermionic solutions are also possible, but their density is subject to the PEP restriction. The fundamental difference is that boson creation operators can be applied to the vacuum to an arbitrary degree, with occupation numbers limited only by thermodynamic equilibrium. For fermions, the degree is bounded above by unity due to the nilpotency of the creation operators, which limits the occupation numbers also for non-equilibrium states.

\begin{figure}
\begin{center}
\parbox{0.4\textwidth}{\includegraphics[width=0.4\textwidth]{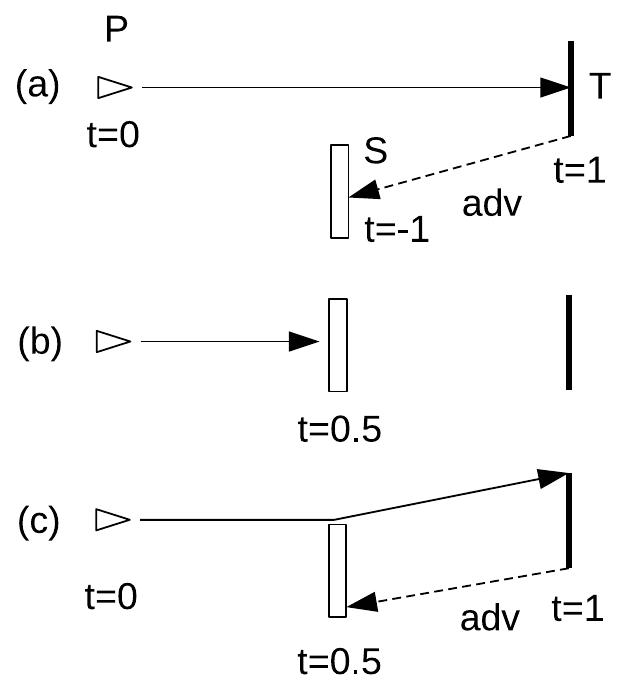}}~~~~~
\parbox{0.5\textwidth}{\caption{Schematic representation of Wheeler-Feynman time machine.}\label{f4}}
\end{center}
\end{figure}

\section{Temporal aspects of the model}
\paragraph{B1: Time loops and causality principle.} This section follows the presentation of \cite{Visser1996}. A space-time implements a {\it time machine} if it contains a closed time-like loop. Examples of solutions include the simplest cylindrical spacetime with cyclic time $S^1\times R^3$, examples with rapid rotation of massive bodies \cite{vanStockum1937,Goedel1949,Kerr1963,Gott1991}, wormholes \cite{MorrisThorne1988}, a recent development is the solution \cite{1310.7985} with a warp-drive \cite{gr-qc/0009013} closed in a loop. The ordinary, linear causality principle, which requires effects to be in the future of causes, does not work for cyclic time. Indeed, for a given event $A$, all events in the time loop, excluding $A$, belong both to the future and to the past of $A$. There is Novikov's conjecture on chronological self-consistency, briefly {\it Novikov's chronology}, which can be considered as a cyclic version of the causality principle. According to this principle, the only requirement for a solution is that the events in the time loop be logically self-consistent.

Violations of this principle lead to contradictions known as {\it causal paradoxes}. In reality, there is only one fundamental contradiction, the so-called grandfather paradox, a situation where $A$ implies $B$, and $B$ implies the logical negation of $A$. It reduces to the well-known liar paradox in logic, $A=(not)A$. There is no solution in this situation. Another situation, when $A$ implies $B$, and $B$ implies $A$, is sometimes called the bootstrap paradox, although in reality it is not a contradiction, but represents the normal course of events in Novikov's chronology.

Although the described examples are constructed for topologically non-trivial space-time in the general theory of relativity, the consideration can be transferred to Minkowski space-time by gluing together time loops from several world lines. An orientation, an arrow of time, must be introduced on the world lines. This may be a thermodynamic arrow, in the direction of increasing entropy, or an information arrow, in the direction of signal propagation. The consideration can also be extended to piecewise light-like curves located in the closure of the set of piecewise time-like ones.

Some aspects of non-trivial chronology were considered in the works of Wheeler-Feynman \cite{wf1945,wf1949}. These works consider T-symmetric version of electrodynamics with retarded and advanced interactions. If the system of charges is surrounded by an absolute absorber, then this model becomes equivalent to standard electrodynamics with radiative friction and delayed interaction. However, the absolute absorption condition is itself T-symmetric, so for the same trajectories there is a T-conjugate description with radiative friction of the opposite sign and advanced interaction. One description can be preferred over another by introducing a thermodynamic arrow of time in the absorber.

The work \cite{wf1949} also built a time machine, schematically depicted in Fig.\ref{f4}a-c. Pellet P hits target T, which sends a signal back in time using an advanced electromagnetic wave. In the past, shutter S closes, which prevents the pellet from hitting the target. Here a time loop (a,b) is constructed with a logically inconsistent solution. The contradiction exists only for the discrete states of the target (hit or not hit) and the shutter (open or closed). Considering a continuous class of trajectories, one can find an intermediate solution (c), in which the shutter is half-open and the pellet hits the target edge. Thus, a self-consistent solution can be constructed that satisfies Novikov's chronology.

In this work there is also a time loop formed by the backward in cosmological time DM flow and the forward LM flow. The most of the processes here are equilibrium, occurring at constant entropy. The arrow of time is available only at the beginning and the end of the processes. From the point of view of an observer related with DM flow, DM is emitted from QBHs and absorbed in BB. From the point of view of an observer related with LM flow, the thermodynamically inverse processes of the separation of CDM will first be observed, and finally DM flows radially converging on QBHs will be observed, while the evolution of LM itself proceeds along its arrow of time. Since there are no logical contradictions along the time loop, the processes satisfy Novikov's chronology.

\paragraph{B2: Self-regulation of DM mass and lifetime of the universe.} In the case of DM dominance, the solution of the Friedman equation $\dot a^2=C_m/a-k$ on the cosmological scale factor $a$ has the form of a cycloid with a maximum size $a_{max }=C_m/k$ and the lifetime of the universe $t_{uni}=C_m\pi/k^{3/2}$. Here the curvature parameter $k>0$ and $C_m=8\pi G\rho_0a_0^3/3$ are proportional to the conserved DM mass. Reexpressing $k$ through the Hubble parameter $H_0=\dot a_0/a_0$, obtain $t_{uni}=C_m\pi a_0^{3/2}(C_m-C_{m,crit})^{-3/2 }$, $C_{m,crit}=a_0^3H_0^2$. Here the moment of time $t_0$, not necessarily today, is chosen in the region of DM dominance. Considering solutions with $C_m>C_{m,crit}$, which is equivalent to $\rho>\rho_{crit}$, with fixed $(a_0,H_0)$, one sees that $t_{uni}$ is a decreasing function of $C_m$. This behavior is reasonable; as DM mass increases, the lifetime of the universe decreases due to gravitational collapse. Another relation $C_m=c_1t_{uni}$ connects the mass of DM to its emission from QBHs. The solution of this system is unique and has the form $t_{uni}=C_m/c_1$, $C_m=C_{m,crit}+a_0c_1^{2/3}\pi^{2/3}$. For $c_1\to0$ obtain $t_{uni}\sim C_{m,crit}/c_1\to\infty$, the solution tends to the flat case. In the general case, one can leave $(a_0,H_0)$ not fixed, assume them to be functions of $c_1$. In this case, it turns out that $k=(C_m-C_{m,crit})/a_0=c_1^{2/3}\pi^{2/3}$, for $c_1\to0$ one obtains $k\to0$, the flat case.

\paragraph{B3: T-asymmetric solutions.} In addition to T-symmetric quasi-black holes considered in this work, solutions in the form of ordinary, T-asymmetric black holes are possible. In the considered cosmological scenario, in backward evolution, such black holes emit DM, thereby representing white holes. Matter is emitted isotropically, reaching large distances, in particular, it can escape to infinity. In forward evolution, such solutions represent black holes continuously absorbing matter, described by well-known gravitational collapse models \cite{gr-qc/0502040,Blau2022}. At the same time, for T-asymmetric solutions, one can adjust the coefficient describing the density of matter flows to the same value as for T-symmetric ones. Thus, the same conditions near BB and the same global characteristics will be obtained, in particular, the same conditions for self-regulation of DM mass. In detail, however, in the distant future, a different evolution of galaxies may result, since in this case, in the end, all DM will be absorbed by the black holes, and a strong redistribution of masses in galaxies may occur.

\paragraph{B4: Cosmological redshift and time reversibility.} When considering the effect of cosmological redshift on massive bodies, the following problem arises. Let the body initially move in a relativistic mode, with an energy significantly exceeding the rest energy, $E\gg mc^2$. The redshift gradually decreases the energy, and as soon as $E=mc^2$, the body stops. If one considers different initial energies, the stop will occur at different times, but the final state will be the same, the state of rest. This is in apparent contradiction with the reversibility of time, since it is unknown what final energy the body will receive during backward evolution from the state of rest.

The problem is resolved by considering timelike geodesics in FLRW metric \cite{Blau2022}. Proper velocity relative to comoving observers, the so-called peculiar velocity is $v/c=(1+a^2/\beta^2)^{-1/2}$, where $\beta$ is constant on the trajectory. Note that this is a formal expression for the velocity, which assumes the universe to be homogeneous and does not take into account gravitational motion after the formation of structures. To analyze the irreversibility problem, such an expression is sufficient. With $a_0=1$ today and choosing $\beta\ll1$, get $v_0/c=\beta$. The mode switches at $a\sim\beta$, so at $a\gg\beta$ the velocities are non-relativistic $v/c=\beta/a$, at $a\ll\beta$ the velocities are relativistic $v/c\sim1$. For example, for the switching moment in the range $a=1.7\cdot10^{-10}-2.5\cdot10^{-9}$, corresponding to the region between ND and BBN, one obtains today speeds $v_0=0.05-0.75m/s$. Thus, for different initial energies the velocities in the final state are different, although they are close to zero in the considered setup. In practice, this means that a bundle of trajectories corresponding to different initial energies can be taken into consideration together with the central trajectory describing the final state of rest in comoving coordinates.

\section{Photon case closed}

This section examines the possibility that DM is formed by the photons of the standard model. Such photons may have a frequency so low that they cannot be detected by conventional means, and a density so high as to reproduce the gravitational effects of DM. In a cosmological context, this possibility is excluded, since in a homogeneous medium, before the formation of structures, the photon gas forms a relativistic DM, which contradicts the observations. Restrictions due to the interaction of photons with ISM/IGM are discussed below, which also rule out the use of photons to describe DM.

\paragraph{C1: Interaction of low frequency EM-waves with ISM/IGM.} Propagation of EM-waves in dilute plasma of ISM/IGM is characterized by plasma frequency $\omega_p(ISM)\sim10^4s^{-1}$, $ \omega_p(IGM)\sim10^2s^{-1}$ \cite{BlandfordThorne}. EM-waves propagate in the medium at a frequency $\omega>\omega_p$, in particular, at $\omega>\omega_p(ISM)$ waves from the central region of the galaxy reach the solar system. In order for such waves to be used in DM models, their intensity must be high enough. For the local DM density used in this work in the vicinity of the Sun, the irradiance is $10~kW/m^2$, compare with the typical Solar irradiance $1.4~kW/m^2$ on Earth orbit \cite{solar}. Radiation of such intensity, emitted mainly from the galactic center, would undoubtedly be detected by ground-based or orbital radio telescopes. The absence of a signal excludes this frequency range. Further, for $\omega_p(IGM)<\omega<\omega_p(ISM)$ the waves do not reach the Sun, but can freely propagate in IGM for sources located in the near-surface zone of the galactic disk. This possibility is also excluded; in the considered DM model, the radiation should not leave the galaxy, but should be encapsulated into a sphere of radius $R_{cut}$, representing an outer halo radius. Only in this case DM appears to be localized in galaxies, and the observed cosmological evolution turns out to be equivalent to CDM model. The remaining case, $\omega<\omega_p(IGM)$, can theoretically reproduce such encapsulation due to a large skin effect. At frequencies less than $\omega_p$, waves do not propagate freely in the medium, but penetrate into it to a distance of the order of skin depth. This depth increases with decreasing frequency, as $\sim\omega^{-1/2}$ for normal and as $\sim\omega^{-1/3}$ for anomalous skin effect. Theoretically, there is a possibility that for ultra-low frequencies the skin depth can reach the size of the galaxy and reproduce the required cutoff at the radius $R_{cut}$. This scenario will be discussed in detail below based on established models of skin effect \cite{KolobovEconomou,Weibel}. In addition, the effect of EM wave pressure on ISM/IGM will be considered.

\begin{figure}
\begin{center}
\parbox{0.4\textwidth}{\includegraphics[width=0.4\textwidth]{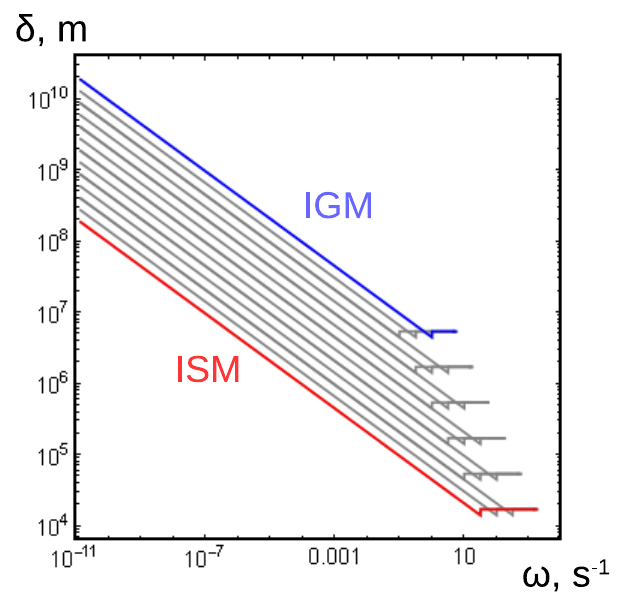}}~~~~~
\parbox{0.5\textwidth}{\caption{Dependence of skin depth on frequency for propagation of EM-waves in ISM/IGM.}\label{f5}}
\end{center}
\end{figure}

\paragraph{C2: Skin effect.} Combining the formulas from \cite{BlandfordThorne,KolobovEconomou}, converting all of them into the SI system, get:
\begin{eqnarray}
&&\omega_p=(n_ee^2/(\epsilon_0m_e))^{1/2},\ v_e=(2k_BT_e/m_e)^{1/2},\\ 
&&\nu_e=(n_ee^4\log\Lambda_C)/(2\pi\epsilon_0^2m_e^2(3/2)^{3/2}v_e^3),\ l_e=v_e/\nu_e,\\
&&\omega_0=4\pi c/R_\odot,\ \omega_1=\omega_p/10,\ \omega_0\leq\omega\leq\omega_1,\\
&&\delta_0=(c/\omega_p)(1+\nu_e^2/\omega^2)^{1/4},\ \epsilon=\arctan(\nu_e/\omega),\\
&&\delta_c=\delta_0/\cos(\epsilon/2),\ \delta_a=(c/\pi^{1/6})((v_e/c)/(\omega\omega_p^2))^{1/3},\\
&&\delta=\left\{\begin{array}{ll}
\delta_c,&(\omega<\nu_e\,\&\,l_e/\delta_0<1)\,|\,(\omega\geq\nu_e\,\&\,l_e/\delta_0<\omega/\nu_e)\\
\delta_a,&\mbox{otherwise}\\
\end{array}\right.
\end{eqnarray}
The parameters used are: fundamental -- elementary charge $e$, electron mass $m_e$, vacuum permittivity $\epsilon_0$, speed of light in vacuum $c$, Boltzmann constant $k_B$, model -- plasma frequency $\omega_p$, electron number density $n_e$, electron thermal velocity $v_e$, electron temperature $T_e$, electron collision frequency $\nu_e$, Coulomb logarithm $\log\Lambda_C$, taken value $\log\Lambda_C=10$, electron mean free path $l_e$, distance from the Sun to the galactic center $R_\odot$, frequency of electromagnetic waves $\omega$, with limits $\omega_{0,1}$, angular parameter $\epsilon$, classical skin depth $\delta_c$, anomalous skin depth $\delta_a$, combined skin depth $\delta$.

Let us use the parameters from \cite{BlandfordThorne}:
\begin{eqnarray}
&&n_e(ISM)=10^5m^{-3},\ T_e(ISM)=10^4K,\label{neISM}\\
&&n_e(IGM)=1m^{-3},\ T_e(IGM)=10^6K.\label{neIGM}
\end{eqnarray}
The upper frequency limit $\omega_1$ is chosen to satisfy the requirement $\omega\ll\omega_p$, at which the used skin effect model is valid. The lower limit $\omega_0$ is found from the condition $R_\odot=2\lambda$, where $\lambda$ is the radiation wavelength, which corresponds to the far-field regime in the region of the solar system. Namely for this regime the amplitudes of EM-fields decrease as $\sim r^{-1}$, and the energy density as $\sim r^{-2}$, which corresponds to the observed flat profile of RC in the vicinity of the solar system.

Fig.\ref{f5} shows $\delta(\omega)$ profiles found using these formulas for ISM/IGM, as well as in gray for intermediate values of $(n_e,T_e)$. Near the upper limit, the classic skin effect is realized, and in most of the graph -- the anomalous one. The slight jump between these modes occurs because the formulas used are estimates and their continuous stitching has not been ensured.

From this graph it follows that the upper limit is $\delta<10^{11}m$, which is significantly less than the sizes of galactic structures $R_\odot\sim10^{20}m$, $R_{cut}\sim10^{21}m$. Even if one allows frequencies below the selected $\omega_0$-boundary, the equality $\delta=R_{cut}$ is achieved at $\omega\sim10^{-44}s^{-1}$, which corresponds to the wavelengths $\lambda \sim10^{52}m$, significantly larger than the size of the observable universe. In summary, for a physically reasonable frequency range, the skin effect occurs in a thin layer compared to the galactic size. Outside this layer, EM-fields are negligible, ruling out the use of such fields to describe DM. More detailed skin effect modeling carried out in \cite{Weibel} leads to the same results.

\paragraph{C3: Pressure of EM-waves.} At the energy density required to describe gravitational DM effects, the pressure turns out to be quite large. The following simple evaluation, carried out in geometric units, shows that this pressure is significantly higher than ISM/IGM self-pressure:
\begin{eqnarray}
&&p_1=\rho_1\gg\rho_2\gg p_2.
\end{eqnarray}
Here $p_1$ and $\rho_1$ are the pressure and the mass density of EM-waves, which represent DM in the considered version of the model, $p_2$ and $\rho_2$ are the pressure and the mass density of ISM/IGM. The first equality in this chain follows from the radial direction of EM-waves, the next inequality from the fact that ISM/IGM density is not enough to describe DM, the next inequality from the non-relativistic nature of ISM/IGM. As a consequence of this estimate, at the density required to describe DM, the pressure of EM-waves will blow ISM out of the galaxy. This additional consideration also rules out the use of EM-waves to describe DM.


\begin{thebibliography}{99}

\bibitem{icm2ps2020}
I.~Nikitin, On dark stars, galactic rotation curves and fast radio bursts, Proc. of the 9th Int. Conf. on Mathematical Modeling in Physical Sciences, IC-MSQUARE 2020, J. Phys.: Conf. Ser. 1730, 012073 (2021); doi:10.1088/1742-6596/1730/1/012073.

\bibitem{1612.04889}
B.~Holdom and J.~Ren, Not quite a black hole, Phys. Rev. D95, 084034 (2017); arXiv:1612.04889.
\bibitem{0902.0346}
M.~Visser, C.~Barceló, S.~Liberati, and S.~Sonego, Small, dark, and heavy: But is it a black hole?, Proc. of the Workshop Black Holes in General Relativity and String Theory, Proc. of Science 075, 010 (2008); arXiv:0902.0346.
\bibitem{1401.6562}
C.~Rovelli and F.~Vidotto, Planck stars, Int. J. Mod. Phys. D23, 1442026 (2014); arXiv:1401.6562.
\bibitem{1409.1501}
C.~Barceló, R.~Carballo-Rubio, L.~J.~Garay, and G.~Jannes, The lifetime problem of evaporating black holes: mutiny or resignation, Class. Quant. Grav. 32, 035012 (2015); arXiv:1409.1501.

\bibitem{cond-mat/9911101}
L.~S.~Schulman, Opposite thermodynamic arrows of time, Phys. Rev. Lett. 83, 5419-5422 (1999); arXiv:cond-mat/9911101.
\bibitem{1803.08930}
L.~Boyle, K.~Finn, and N.~Turok, The Big Bang, CPT, and neutrino dark matter, Ann. of Phys. 438, 168767 (2022); arXiv:1803.08930.

\bibitem{9502091}
M.~Persic and P.~Salucci, Rotation curves of 967 spiral galaxies, Astrophys. J. Supp. 99, 501 (1995); arXiv:astro-ph/9502091.
\bibitem{9503051} 
M.~Persic, P.~Salucci, and F.~Stel, Rotation curves of 967 spiral galaxies: Implications for dark matter, Astrophys. Lett. Comm. 33, 205-211 (1996); arXiv:astro-ph/9503051.
\bibitem{9506004} 
M.~Persic, P.~Salucci, and F.~Stel, The universal rotation curve of spiral galaxies: I. The dark matter connection, MNRAS 281, 27-47 (1996); arXiv:astro-ph/9506004.
\bibitem{0703115}
P.~Salucci, A.~Lapi, C.~Tonini, G.~Gentile, I.~Yegorova, and U.~Klein, The universal rotation curve of spiral galaxies: II. The dark matter distribution out to the virial radius, MNRAS 378, 41-47 (2007); arXiv:astro-ph/0703115.
\bibitem{1609.06903}
E.~V.~Karukes and P.~Salucci, The universal rotation curve of dwarf disk galaxies, MNRAS 465, 4703-4722 (2017); arXiv:1609.06903.
\bibitem{0811.0859} 
Y.~Sofue, M.~Honma, and T.~Omodaka, Unified rotation curve of the Galaxy -- decomposition into de Vaucouleurs bulge, disk, dark halo, and the 9-kpc rotation dip, Pub. Astron. Soc. Jap. 61, 227-236 (2009); arXiv:0811.0859.
\bibitem{0811.0860} 
Y.~Sofue, Pseudo rotation curve connecting the Galaxy, dark halo, and Local Group, Pub. Astron. Soc. Jap. 61, 153-161 (2009); arXiv:0811.0860.
\bibitem{1110.4431} 
Y.~Sofue, A grand rotation curve and dark matter halo in the Milky Way Galaxy, Pub. Astron. Soc. Jap. 64, 75 (2012); arXiv:1110.4431.
\bibitem{1307.8241} 
Y.~Sofue, Rotation curve and mass distribution in the Galactic Center -- from black hole to entire Galaxy, Pub. Astron. Soc. Jap. 65, 118 (2013); arXiv:1307.8241.

\bibitem{NFW} 
J.~F.~Navarro, C.~S.~Frenk, and S.~D.~M.~White, The structure of cold dark matter halos, Astrophys. J. 462, 563-575 (1996); arXiv:astro-ph/9508025.
\bibitem{Burkert} 
A.~Burkert, The structure of dark matter halos in dwarf galaxies, Astrophys. J. 447, L25 (1995); arXiv:astro-ph/9504041.
\bibitem{Einasto} 
J.~Einasto, in: Kinematics and Dynamics of Stellar Systems, Alma-Ata, p.87 (1965).

\bibitem{wf1945} 
J.~A.~Wheeler and R.~P.~Feynman, Interaction with the absorber as the mechanism of radiation, Rev. Mod. Phys. 17, 157 (1945).
\bibitem{wf1949} 
J.~A.~Wheeler and R.~P.~Feynman, Classical electrodynamics in terms of direct interparticle action, Rev. Mod. Phys. 21, 425 (1949).

\bibitem{vanStockum1937}
W.~J.~van Stockum, Gravitational field of a distrubution of particles rotating about an axis of symmetry, Proc. R. Soc. Edin. 57, 135-154 (1937). 
\bibitem{Goedel1949}
K.~Gödel, An example of a new type of cosmological solution of Einstein's field equation of gravitation, Rev. Mod. Phys. 21, 447-450 (1949).
\bibitem{Kerr1963}
S.~W.~Hawking and G.~F.~R.~Ellis, The Large Scale Structure of Space-Time, Cambrigde University Press 1973, p.162.
\bibitem{Gott1991}
J.~R.~Gott~III, Closed timelike curves produced by pairs of moving cosmic strings: Exact solutions, Phys. Rev. Lett. 66, 1126-1129 (1991).
\bibitem{MorrisThorne1988}
M.~S.~Morris, K.~S.~Thorne, and U.~Yurtsever, Wormholes, time machines, and the weak energy condition, Phys. Rev. Lett. 61, 1446-1449 (1988).

\bibitem{1310.7985}
B.~K.~Tippett and D.~Tsang, Traversable acausal retrograde domains in spacetime, Class. Quant. Grav. 34, 095006 (2017); arXiv:1310.7985.
\bibitem{gr-qc/0009013}
M.~Alcubierre, The warp drive: hyper-fast travel within general relativity, Class. Quant. Grav., 11 (5), L73-L77 (1994); arXiv:gr-qc/0009013.
\bibitem{Visser1996}
M.~Visser, Lorentzian Wormholes: from Einstein to Hawking, Springer 1996.

\bibitem{BlandfordThorne}
R.~D.~Blandford and K.~S.~Thorne, Applications of Classical Physics, Caltech PMA 2012;
www.pmaweb.caltech.edu/Courses/ph136/yr2012

\bibitem{solar}
C.~A.~Gueymard, A reevaluation of the solar constant based on a 42-year total solar irradiance time series and a reconciliation of spaceborne observations, Solar Energy 168, 2-9 (2018).

\bibitem{KolobovEconomou}
V.~I.~Kolobov and D.~J.~Economou, The anomalous skin effect in gas discharge plasmas, Plasma Sources Sci. Technol. 6, R1-R17 (1997).

\bibitem{Weibel}
E.~S.~Weibel, Anomalous skin effect in a plasma, Physics of Fluids 10, 741 (1967).

\bibitem{gr-qc/0411062} 
A.~J.~S.~Hamilton and S.~E.~Pollack, Inside charged black holes: II. Baryons plus dark matter, Phys. Rev. D 71, 084032 (2005); arXiv:gr-qc/0411062.

\bibitem{freeman} 
K.~C.~Freeman, On the disks of spiral and S0 galaxies, Astrophys. J. 160, 811 (1970).

\bibitem{KATRIN2022} 
M.~Aker et al. (KATRIN Collaboration), Direct neutrino-mass measurement with sub-electronvolt sensitivity, Nature Physics 18, 160-166 (2022).

\bibitem{2106.15267} 
E.~Di~Valentino, S.~Gariazzo, and O.~Mena, On the most constraining cosmological neutrino mass bounds, Phys. Rev. D 104, 083504 (2021); arXiv:2106.15267.

\bibitem{1003.3101} 
P.~Salucci, F.~Nesti, G.~Gentile, and C.~Frigerio~Martins, The dark matter density at the Sun's location, Astronomy and Astrophysics 523, A83 (2010); arXiv:1003.3101. 

\bibitem{2012.11477} 
P.~F.~de~Salas and A.~Widmark, Dark matter local density determination: recent observations and future prospects, Rep. Prog. Phys. 84, 104901 (2021); arXiv:2012.11477.

\bibitem{hep-ph/0501066}
A.~D.~Dolgov and A.~Yu.~Smirnov, Possible violation of the spin-statistics relation for neutrinos: cosmological and astrophysical consequences, Phys. Lett. B621, 1-10 (2005); arXiv:hep-ph/0501066.

\bibitem{0911.5012}
J.~R.~Bhatt, B.~R.~Desai, E.~Ma, G.~Rajasekaran, and U.~Sarkar, Neutrino condensate as origin of dark energy, Phys. Lett. B687, 75-78 (2010); arXiv:0911.5012.
\bibitem{1008.5214}
M.~Azam, J.~R.~Bhatt, and U.~Sarkar, Experimental signatures of cosmological neutrino condensation, Phys. Lett. B697, 7-10 (2011); arXiv:1008.5214.
\bibitem{2004.03731}
A.~Chodos and F.~Cooper, Neutrino condensation from a new Higgs interaction, Phys. Rev. D 102, 113003 (2020); arXiv:2004.03731.

\bibitem{Chavanis1}
P.-H.~Chavanis, The maximum mass of dilute axion stars, The Sixteenth Marcel Grossmann Meeting, 2149-2173 (2023), World Scientific.
\bibitem{Chavanis2}
P.-H.~Chavanis, The self-gravitating Fermi gas in Newtonian gravity and general relativity, The Sixteenth Marcel Grossmann Meeting, 2230-2251 (2023), World Scientific; arXiv:2112.02654.

\bibitem{2202.04587}
M.~Aker et al. (KATRIN Collaboration), New constraint on the local relic neutrino background overdensity with the first KATRIN data runs, Phys. Rev. Lett. 129, 011806 (2022); arXiv:2202.04587.
\bibitem{2207.02860}
V.~Brdar, P.~S.~B.~Dev, R.~Plestid, and A.~Soni, A new probe of relic neutrino clustering using cosmogenic neutrinos, Phys. Lett. B 833, 137358 (2022); arXiv:2207.02860.
\bibitem{2111.14870}
J.~Alvey, M.~Escudero, N.~Sabti, and T.~Schwetz, Cosmic neutrino background detection in large-neutrino-mass cosmologies, Phys. Rev. D 105, 063501 (2022); arXiv:2111.14870.

\bibitem{Blau2022}
M.~Blau, Lecture Notes on General Relativity, University of Bern 2022;\\ www.blau.itp.unibe.ch/GRLecturenotes.html
\bibitem{Tong2019}
D.~Tong, Lectures on Cosmology, University of Cambridge 2019;\\ www.damtp.cam.ac.uk/user/tong/cosmo.html

\bibitem{1808.02472}
A.~Banihashemi, N.~Khosravi, and A.~H.~Shirazi, Ups and downs in dark energy: phase transition in dark sector as a proposal to lessen cosmological tensions, Phys. Rev. D101, 123521 (2020); arXiv:1808.02472.
\bibitem{2012.01407}
A.~Banihashemi, N.~Khosravi, and A.~Shafieloo, Dark energy as a critical phenomenon: a hint from Hubble tension, J. of Cosmology and Astroparticle Phys. 2021, 003 (2021); arXiv:2012.01407.
\bibitem{1809.05678}
A.~B.~Balakin and A.~S.~Ilin, Dark energy and dark matter interaction: kernels of Volterra type and coincidence problem, Symmetry 10, 411 (2018); arXiv:1809.05678.
\bibitem{1812.06854}
W.~Yang, N.~Banerjee, A.~Paliathanasis, and S.~Pan, Reconstructing the dark matter and dark energy interaction scenarios from observations, Phys. Dark Univ. 26, 100383 (2019); arXiv:1812.06854.
\bibitem{1911.04520}
G.~Cheng, Y.-Z.~Ma, F.~Wu, J.~Zhang, and X.~Chen, Testing interacting dark matter and dark energy model with cosmological data, Phys. Rev. D102, 043517 (2020); arXiv:1911.04520.
\bibitem{1903.02370}
A.~Paliathanasis, S.~Pan, and W.~Yang, Dynamics of nonlinear interacting dark energy models, Int. J. Mod. Phys. D28, 1950161 (2019); arXiv:1903.02370.
\bibitem{2010.10823}
Z.~Rezaei, Dark matter -- dark energy interaction and the shape of cosmic voids, The Astrophys. J. 902, 102 (2020); arXiv:2010.10823.
\bibitem{2002.06127}
M.~Lucca and D.~C.~Hooper, Tensions in the dark: shedding light on dark matter -- dark energy interactions, Phys. Rev. D102, 123502 (2020); arXiv:2002.06127.
\bibitem{1804.08558}
W.~Yang, S.~Pan, and A.~Paliathanasis, Cosmological constraints on an exponential interaction in the dark sector, Mon. Not. Roy. Astron. Soc. 482, 1007-1016 (2019); arXiv:1804.08558.
\bibitem{1812.03540}
V.~H.~Cárdenas, D.~Grandón, and S.~Lepe, Dark energy and dark matter interaction in light of the second law of thermodynamics, The European Phys. J. C79, 357 (2019); arXiv:1812.03540.
\bibitem{1907.12551}
S.~Pan, W.~Yang, E.~Di~Valentino, A.~Shafieloo, and S.~Chakraborty, Reconciling $H_0$ tension in a six parameter space?, J. of Cosmology and Astroparticle Phys. 2020, 062 (2020); arXiv:1907.12551.
\bibitem{2001.05103}
X.~Li and A.~Shafieloo, Evidence for emergent dark energy, The Astrophys. J. 902, 58 (2020); arXiv:2001.05103.
\bibitem{1908.04281}
E.~Di~Valentino, A.~Melchiorri, O.~Mena, and S.~Vagnozzi, Interacting dark energy in the early 2020s: a promising solution to the $H_0$ and cosmic shear tensions, Phys. Dark Univ. 30, 100666 (2020); arXiv:1908.04281.
\bibitem{1910.09853}
E.~Di~Valentino, A.~Melchiorri, O.~Mena, and S.~Vagnozzi, Non-minimal dark sector physics and cosmological tensions, Phys. Rev. D101, 063502 (2020); arXiv:1910.09853.
\bibitem{2009.12620}
E.~Di~Valentino and O.~Mena, A fake interacting dark energy detection?, Mon. Not. Roy. Astron. Soc. Lett. 500, L22-L26 (2021); arXiv:2009.12620.
\bibitem{1902.10636}
K.~L.~Pandey, T.~Karwal, and S.~Das, Alleviating the $H_0$ and $\sigma_8$ anomalies with a decaying dark matter model, J. of Cosmology and Astroparticle Phys. 2020, 026 (2020); arXiv:1902.10636.
\bibitem{1908.09843}
S.~Ghosh, R.~Khatri, and T.~S.~Roy, Dark neutrino interactions phase out the Hubble tension, Phys. Rev. D102, 123544 (2020); arXiv:1908.09843.
\bibitem{2002.03408}
S.~Pan, W.~Yang, and A.~Paliathanasis, Nonlinear interacting cosmological models after Planck 2018 legacy release and the $H_0$ tension, Mon. Not. Roy. Astron. Soc. 493, 3114-3131 (2020); arXiv:2002.03408.
\bibitem{1908.03324}
S.~Panpanich, P.~Burikham, S.~Ponglertsakul, and L.~Tannukij, Resolving Hubble tension with quintom dark energy model, Chin. Phys. C45, 015108 (2021); arXiv:1908.03324.
\bibitem{1907.01496}
M.~Archidiacono, D.~C.~Hooper, R.~Murgia, S.~Bohr, J.~Lesgourgues, and M.~Viel, Constraining dark matter -- dark radiation interactions with CMB, BAO, and Lyman-$\alpha$, J. of Cosmology and Astroparticle Phys. 2019, 055 (2019); arXiv:1907.01496.
\bibitem{1906.09189}
M.~Martinelli and I.~Tutusaus, CMB Tensions with low-redshift $H_0$ and $S_8$ measurements: impact of a redshift-dependent type-Ia supernovae intrinsic luminosity, Symmetry 11, 986 (2019); arXiv:1906.09189.
\bibitem{1701.08165}
G.-B.~Zhao et al., Dynamical dark energy in light of the latest observations, Nature Astronomy 1, 627-632 (2017); arXiv:1701.08165.
\bibitem{1807.03772}
Y.~Wang, L.~Pogosian, G.-B.~Zhao, and A.~Zucca, Evolution of dark energy reconstructed from the latest observations, The Astrophys. J. Lett. 869, L8 (2018); arXiv:1807.03772.

\bibitem{gr-qc/0211001}
M.~Visser and N.~Yunes, Power laws, scale invariance, and generalized Frobenius series: Applications to Newtonian and TOV stars near criticality, Int. J. Mod. Phys. A18, 3433-3468 (2003); arXiv:gr-qc/0211001.

\bibitem{gr-qc/0502040}
R.~J.~Adler, J.~D.~Bjorken, P.~Chen, and J.~S.~Liu, Simple analytic models of gravitational collapse, Am. J. Phys. 73 1148-1159 (2005); arXiv:gr-qc/0502040.

\end{thebibliography}
\end{document}